\documentclass[conference]{IEEEtran} 
\IEEEoverridecommandlockouts
\usepackage{cite}
\usepackage{amsmath,amssymb,amsfonts}
\usepackage{algorithmic}
\usepackage{graphicx}
\usepackage{textcomp}
\usepackage{xcolor}
\def\BibTeX{{\rm B\kern-.05em{\sc i\kern-.025em b}\kern-.08em
    T\kern-.1667em\lower.7ex\hbox{E}\kern-.125emX}}
\begin{document}

\title{Multi-Temporal Analysis and Scaling Relations of 100,000,000,000 Network Packets
\thanks{This material is based upon work supported by the Assistant Secretary of Defense for Research and Engineering under Air Force Contract No. FA8702-15-D-0001, National Science Foundation CCF-1533644, and United States Air Force Research Laboratory Cooperative Agreement Number FA8750-19-2-1000. Any opinions, findings, conclusions or recommendations expressed in this material are those of the author(s) and do not necessarily reflect the views of the Assistant Secretary of Defense for Research and Engineering, the National Science Foundation, or the United States Air Force. The U.S. Government is authorized to reproduce and distribute reprints for Government purposes notwithstanding any copyright notation herein.}
}

\author{\IEEEauthorblockN{Jeremy Kepner$^{1,2,3}$, Chad Meiners$^{4}$, Chansup Byun$^{1}$, Sarah McGuire$^{4}$, Timothy Davis$^5$,  William Arcand$^{1}$, \\ Jonathan Bernays$^{4}$, David Bestor$^{1}$, William Bergeron$^{1}$, Vijay Gadepally$^{1,2}$, Raul Harnasch$^{4}$,  Matthew Hubbell$^{1}$, \\ Micheal Houle$^{1}$, Micheal Jones$^{1}$, Andrew Kirby$^1$, Anna Klein$^{1}$, Lauren Milechin$^{6}$, Julie Mullen$^{1}$,  Andrew Prout$^{1}$, \\  Albert Reuther$^{1}$, Antonio Rosa$^{1}$, Siddharth Samsi$^{1}$, Doug Stetson$^{4}$, Adam Tse$^{4}$, Charles Yee$^{1}$, Peter Michaleas$^{1}$
\\
\IEEEauthorblockA{$^1$MIT Lincoln Laboratory Supercomputing Center,  $^2$MIT Computer Science \& AI Laboratory, \\$^3$MIT Mathematics Department, $^4$MIT Lincoln Laboratory Cyber Operations \& Analysis Technology Group, \\ $^5$Texas A\&M, $^6$MIT Dept. of Earth, Atmospheric, \& Planetary Sciences
}}}
\maketitle

\begin{abstract}
Our society has never been more dependent on computer networks.  Effective utilization of networks requires a detailed understanding of the normal background behaviors of network traffic.  Large-scale measurements of networks are computationally challenging.  Building on prior work in interactive supercomputing and GraphBLAS hypersparse hierarchical traffic matrices, we have developed an efficient method for computing a wide variety of streaming network quantities on diverse time scales.  Applying these methods to 100,000,000,000 anonymized source-destination pairs collected at a network gateway reveals many previously unobserved scaling relationships.  These observations provide new insights into normal network background traffic that could be used for anomaly detection, AI feature engineering, and testing theoretical models of streaming networks.
\end{abstract}

\begin{IEEEkeywords}
Internet modeling, packet capture, streaming graphs, power-law networks, hypersparse matrices
\end{IEEEkeywords}

\section{Introduction}
\begin{figure}
\center{\includegraphics[width=1.0\columnwidth]{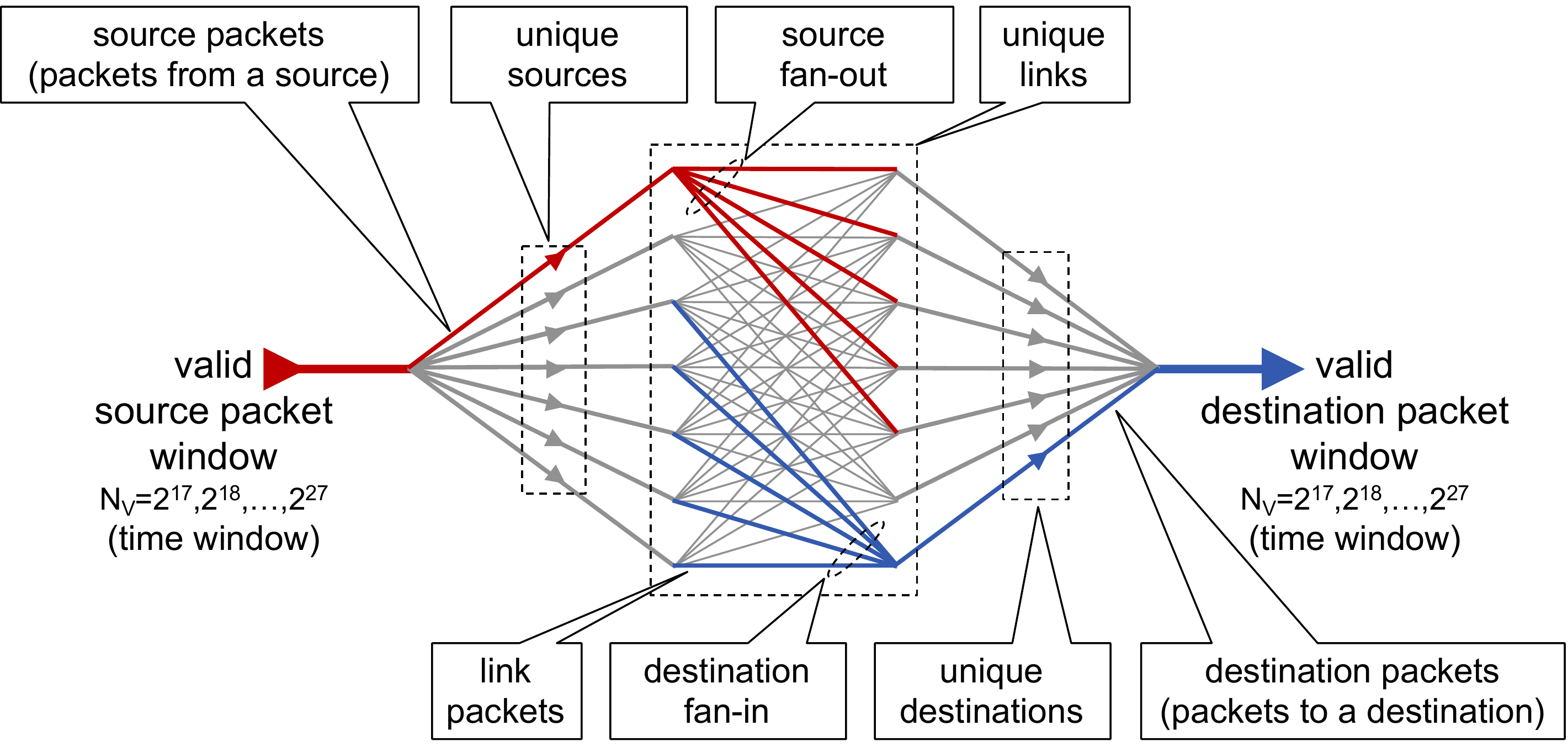}}
      	\caption{{\bf Streaming network traffic quantities.} Internet traffic streams of $N_V$ valid packets are divided into a variety of quantities for analysis: source packets, source fan-out, unique source-destination pair packets (or links), destination fan-in, and destination packets.}
      	\label{fig:NetworkDistribution}
\end{figure}

Global usage of the Internet is expected to exceed 5  billion people in 2023\cite{Cisco2018-2023}.  Accordingly, cyberspace is a frontier as worthy of scientific study as land, sea, air, and space were during past eras of exploration \cite{Bush1945,claffy1999internet,prasad2003bandwidth,kim2008internet,boguna2009navigability}.  Deepening our scientific understanding of cyberspace is expected to yield correspondingly equivalent societal benefits \cite{claffy2000measuring,li2013survey,rabinovich2016measuring}.  More pragmatically, effective utilization of cyberspace requires a detailed understanding of its principal characteristic: network traffic \cite{ClaffyClark2020}.  Figure~\ref{fig:NetworkDistribution} illustrates essential quantities found in all streaming dynamic networks. These quantities are derivable from the source and destinations in packet headers that are the foundational transactional unit of the Internet \cite{huang2018software}. 

Precise measurements of these quantities on networks as vast as the Internet are computationally challenging \cite{lumsdaine2007challenges,kolda2009tensor,hilbert2011world}. Prior work has developed massively parallel data processing capabilities that can effectively use thousands of processors to rapidly process billions of packets in a few hours \cite{Kepner2009,kepner2011graph,kepner2018mathematics,reuther2018interactive,gadepally2018hyperscaling,kepner19streaming}.  Using these capabilities, 50,000,000,000 anonymized packets from the largest publicly available collections of Internet packet data were analyzed, demonstrating the ubiquity of the Zipf-Mandelbrot distribution  and the importance of the size of the packet sampling window $N_V$.  These data consisted of four 1-hour and two 2-day collections from Internet ``trunk'' lines, providing a high-level view of their portions of the Internet spanning years and continents \cite{cho2000tr,borgnat2009seven,dainotti2012issues,himura2013synoptic,fontugne2017scaling,dhamdhere2018inferring}.  These results underscore the need for processing longer time periods from additional vantage points.

Network gateways are a common view into the Internet, are widely monitored, and represent a natural observation point of network traffic.  Using such a gateway we were able to collect trillions of packet headers over several years.  Processing such a large volume of data requires additional computational innovations.  The incorporation of GraphBLAS hypersparse hierarchical traffic matrices has enabled the processing of hundreds of billions of packets in minutes \cite{kepner16mathematical,buluc17design,davis18algorithm,kepner202075}.  This paper presents an initial analysis of one month of anonymized source-destination pairs derived from approximately 100,000,000,000 packets.  The contiguous nature of these data allows the exploration of a wide range of packet windows from $N_V = 2^{17}$ (seconds) to $N_V = 2^{27}$ (hours), providing a unique view into how network quantities depend upon time.  These observations provide new insights into normal network background traffic that could be used for anomaly detection, AI feature engineering, polystore index learning, and testing theoretical models of streaming networks \cite{elmore2015demonstration,kraska18case,do20classifying}.

The outline of the rest of the paper is as follows.  First, the network quantities and their distributions are defined in terms of traffic matrices.   Second, multi-temporal analysis of hypersparse hierarchical traffic matrices is described in the context of network gateways.  Third, the method for determining scaling relations as a function of the packet window $N_V$ is presented along with the resulting scaling relations observed in the gateway traffic data.  Fourth, an analysis of simple network topologies is performed to illustrate that  the  observed relations fall between well-defined models.  Finally, our conclusions and directions for further work are presented.

\section{Network Quantities and Distributions from Traffic Matrices}

The network quantities depicted in Figure~\ref{fig:NetworkDistribution} are computable from origin-destination  matrices that are widely used to represent network traffic \cite{soule2004identify,zhang2005estimating,mucha2010community,tune2013internet}.  It is common to filter the packets down to a valid set for  any particular analysis.   Such filters may limit particular sources, destinations, protocols, and time windows. To reduce statistical fluctuations, the streaming data should be partitioned so that for any chosen time window all data sets have the same number of valid packets \cite{kepner19streaming}.  At a given time $t$, $N_V$ consecutive valid packets are aggregated from the traffic into a sparse matrix ${\bf A}_t$, where ${\bf A}_t(i,j)$ is the number of valid packets between the source $i$ and destination $j$. The sum of all the entries in ${\bf A}_t$ is equal to $N_V$
  \begin{equation}\label{eq:Valid}
    \sum_{i,j} {\bf A}_t(i,j) = N_V
  \end{equation}
All the network quantities depicted in Figure~\ref{fig:NetworkDistribution} can be readily computed from ${\bf A}_t$ using the formulas listed in Table~\ref{tab:Aggregates}.  

\begin{table}
\caption{Network Quantities from Traffic Matrices}
\vspace{-0.25cm}
Formulas for computing network quantities from  traffic matrix ${\bf A}_t$ at time $t$ in both summation and matrix notation. ${\bf 1}$ is a column vector of all 1's, $^{\sf T}$  is the transpose operation, and $|~|_0$ is the zero-norm that sets each nonzero value of its argument to 1\cite{karvanen2003measuring}.
\begin{center}
\begin{tabular}{p{1.45in}p{0.9in}p{0.6in}}
\hline
{\bf Aggregate} & {\bf ~~~~Summation} & {\bf ~Matrix} \\
{\bf Property} & {\bf ~~~~~~Notation} & {\bf Notation} \\
\hline
Valid packets $N_V$ & $~~\sum_i ~ \sum_j ~ {\bf A}_t(i,j)$ & $~{\bf 1}^{\sf T} {\bf A}_t {\bf 1}$ \\
Unique links & $~~\sum_i ~ \sum_j |{\bf A}_t(i,j)|_0$  & ${\bf 1}^{\sf T}|{\bf A}_t|_0 {\bf 1}$ \\
Link packets from $i$ to $j$ & $~~~~~~~~~~~~~~{\bf A}_t(i,j)$ & ~~~$~{\bf A}_t$ \\
Max link packets ($d_{\rm max}$) & $~~~~~\max_{ij}{\bf A}_t(i,j)$ & $\max({\bf A}_t)$ \\
\hline
Unique sources & $~\sum_i |\sum_j ~ {\bf A}_t(i,j)|_0$  & ${\bf 1}^{\sf T}|{\bf A}_t {\bf 1}|_0$ \\
Packets from source $i$ & $~~~~~~~\sum_j ~ {\bf A}_t(i,j)$ & ~~$~~{\bf A}_t  {\bf 1}$ \\
Max source packets ($d_{\rm max}$)  & $ \max_i \sum_j ~ {\bf A}_t(i,j)$ & $\max({\bf A}_t {\bf 1})$ \\
Source fan-out from $i$ & $~~~~~~~~~~\sum_j |{\bf A}_t(i,j)|_0$  & ~~~$|{\bf A}_t|_0 {\bf 1}$ \\
Max source fan-out ($d_{\rm max}$) & $ \max_i \sum_j |{\bf A}_t(i,j)|_0$  & $\max(|{\bf A}_t|_0 {\bf 1})$ \\
\hline
Unique destinations & $~\sum_j |\sum_i ~ {\bf A}_t(i,j)|_0$ & $|{\bf 1}^{\sf T} {\bf A}_t|_0 {\bf 1}$ \\
Destination packets to $j$ & $~~~~~~~\sum_i ~ {\bf A}_t(i,j)$ & ${\bf 1}^{\sf T}|{\bf A}_t|_0$ \\
Max destination packets ($d_{\rm max}$) & $ \max_j \sum_i ~ {\bf A}_t(i,j)$ & $\max({\bf 1}^{\sf T}|{\bf A}_t|_0)$ \\
Destination fan-in to $j$ & $~~~~~~~~~~\sum_i |{\bf A}_t(i,j)|_0$ & ${\bf 1}^{\sf T}~{\bf A}_t$ \\
Max destination fan-in ($d_{\rm max}$) & $ \max_j \sum_i |{\bf A}_t(i,j)|_0$ & $\max({\bf 1}^{\sf T}~{\bf A}_t)$ \\
\hline
\end{tabular}
\end{center}
\label{tab:Aggregates}
\end{table}%

\begin{figure}
\center{\includegraphics[width=0.8\columnwidth]{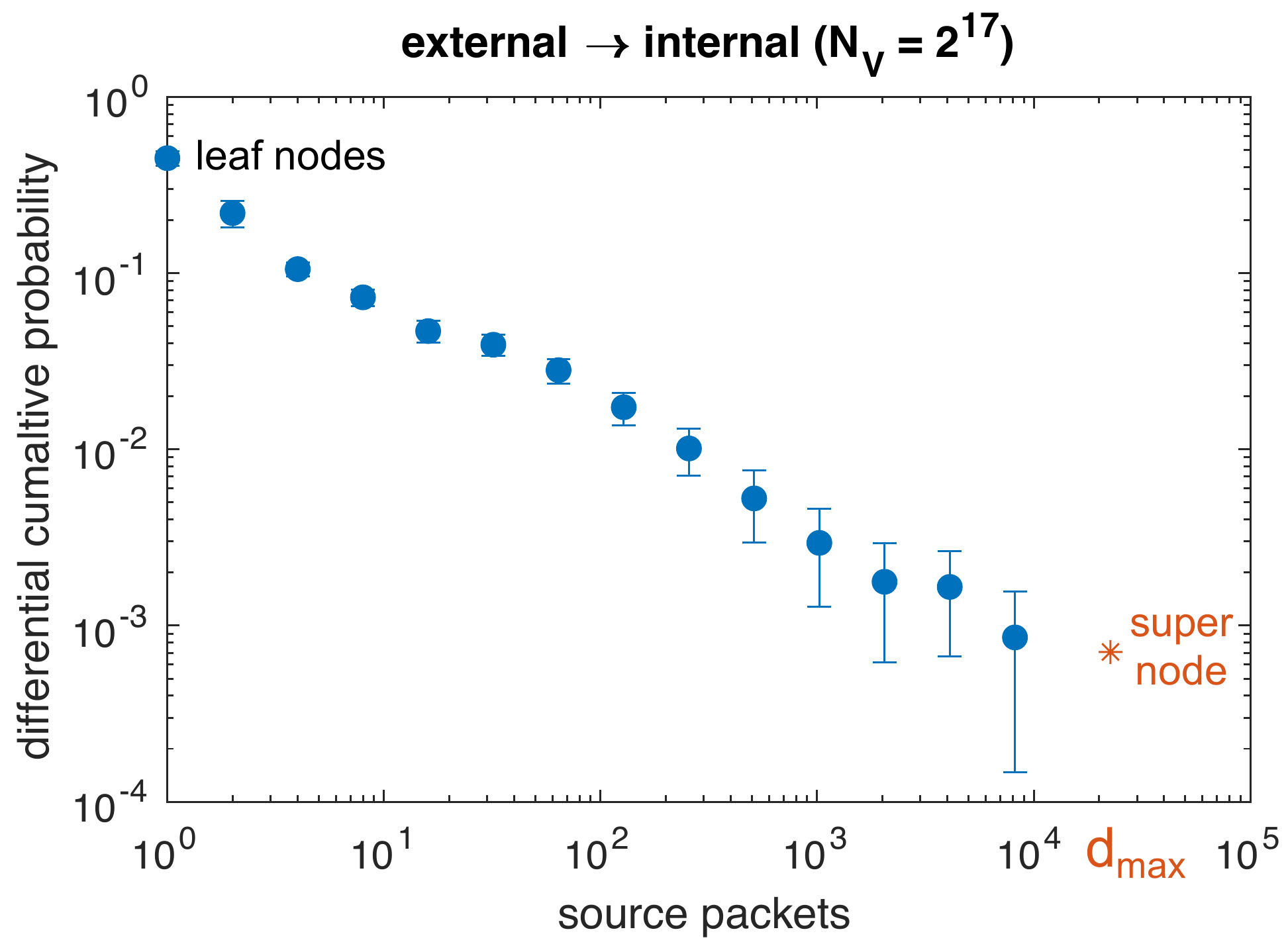}}
      	\caption{{\bf External $\rightarrow$ internal source packet degree distribution.}  Differential cumulative probability (normalized histogram) of the number (degree) of source packets from each source using logarithmic bins $d_i = 2^i$.  Circles represent the averages of  1024 packet windows each with $N_V=2^{17}$.  Error bars represent one standard deviation.  Sources sending out a single packet are denoted ``leaf nodes''.  The source with the largest number of packets $d_{\rm max}$ is referred to as the ``supernode''.}
      	\label{fig:DegreeDistribution}
\end{figure}

Each network quantity will produce a distribution of values whose magnitude is often called the degree $d$. The corresponding histogram of the network quantity is denoted by $n_t(d)$.  The largest observed value in the distribution is denoted  $d_{\rm max}$.  The normalization factor of the distribution is given by
  \begin{equation}\label{eq:Norm}
    \sum_d n_t(d)
  \end{equation}
with corresponding probability
  \begin{equation}\label{eq:Probability}
    p_t(d) = n_t(d)/\sum_d n_t(d)
  \end{equation}
and cumulative probability
  \begin{equation}\label{eq:Cumulative}
    P_t(d) = \sum_{i=1,d} p_t(d)
  \end{equation}
Because of the relatively large values of $d$ observed, the measured probability at large $d$ often exhibits large fluctuations. However, the cumulative probability lacks sufficient detail to see variations around specific values of $d$, so it is typical to pool the \emph{differential cumulative probability} with logarithmic bins in $d$
  \begin{equation}\label{eq:LogBin}
    D_t(d_i) = P_t(d_i) - P_t(d_{i-1})
  \end{equation}
where $d_i = 2^i$ \cite{clauset2009power}.  All computed probability distributions use the same binary logarithmic binning to allow for consistent statistical comparison across data sets \cite{clauset2009power,barabasi2016network}.  The corresponding mean and standard deviation of $D_t(d_i)$ over many different consecutive values of $t$ for a given data set are denoted $D(d_i)$ and $\sigma(d_i)$. Figure~\ref{fig:DegreeDistribution} provides an example distribution of external $\rightarrow$ internal source packets using a packet window of $N_V = 2^{17}$.  The means and standard deviations are computed using 1024  consecutive packet windows.  The resulting distribution exhibits the power-law frequently observed in network  data  \cite{leland1994self,faloutsos1999power,albert1999internet,barabasi1999emergence,adamic2000power,barabasi2009scale,mahanti2013tale}.

\section{Multi-Temporal Analysis}

\begin{figure}
\center{\includegraphics[width=1.0\columnwidth]{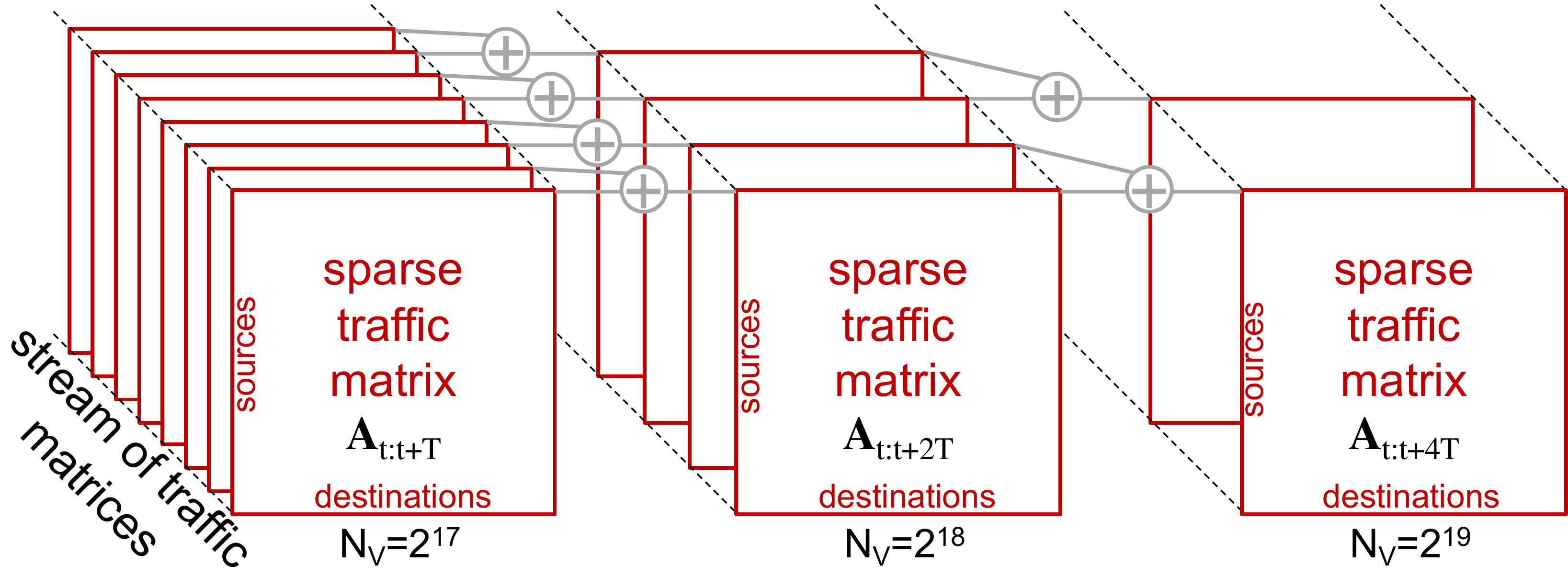}}
      	\caption{{\bf Multi-temporal streaming traffic matrices.} Efficient computation of network quantities on multiple time scales can be achieved by hierarchically aggregating data in different time windows.}
      	\label{fig:MultiTemporalMatrix}
\end{figure}

Network traffic is dynamic and exhibits varying behavior on a wide range of time scales.  A given packet window size $N_V$ will be sensitive to phenomena on its corresponding timescale.  Determining how network quantities scale with $N_V$ provides insight into the temporal behavior of network traffic.   Efficient computation of network quantities on multiple time scales can be achieved by hierarchically aggregating data in different time windows \cite{kepner19streaming}.  Figure~\ref{fig:MultiTemporalMatrix} illustrates a binary aggregation of  different streaming traffic matrices.   Computing each quantity at each hierarchy level eliminates redundant computations that would be performed if each packet window was computed separately.  Hierarchy also ensures that most computations are performed on smaller matrices residing in faster memory.  Correlations among the matrices mean  that adding two matrices each with $N_V$ entries results in a matrix with fewer than $2N_V$ entries, reducing the relative number of operations as the matrices grow.

One of the important capabilities of the SuiteSparse GraphBLAS library is support for hypersparse matrices where the number of nonzero entries is significantly less than either dimensions of the matrix.  If the packet source and destination identifiers are drawn from a large numeric range, such as those used in the Internet protocol, then a hypersparse representation of ${\bf A}_t$ eliminates the need to keep track of additional indices and can significantly accelerate the computations \cite{kepner202075}.

\begin{figure}
\center{\includegraphics[width=0.65\columnwidth]{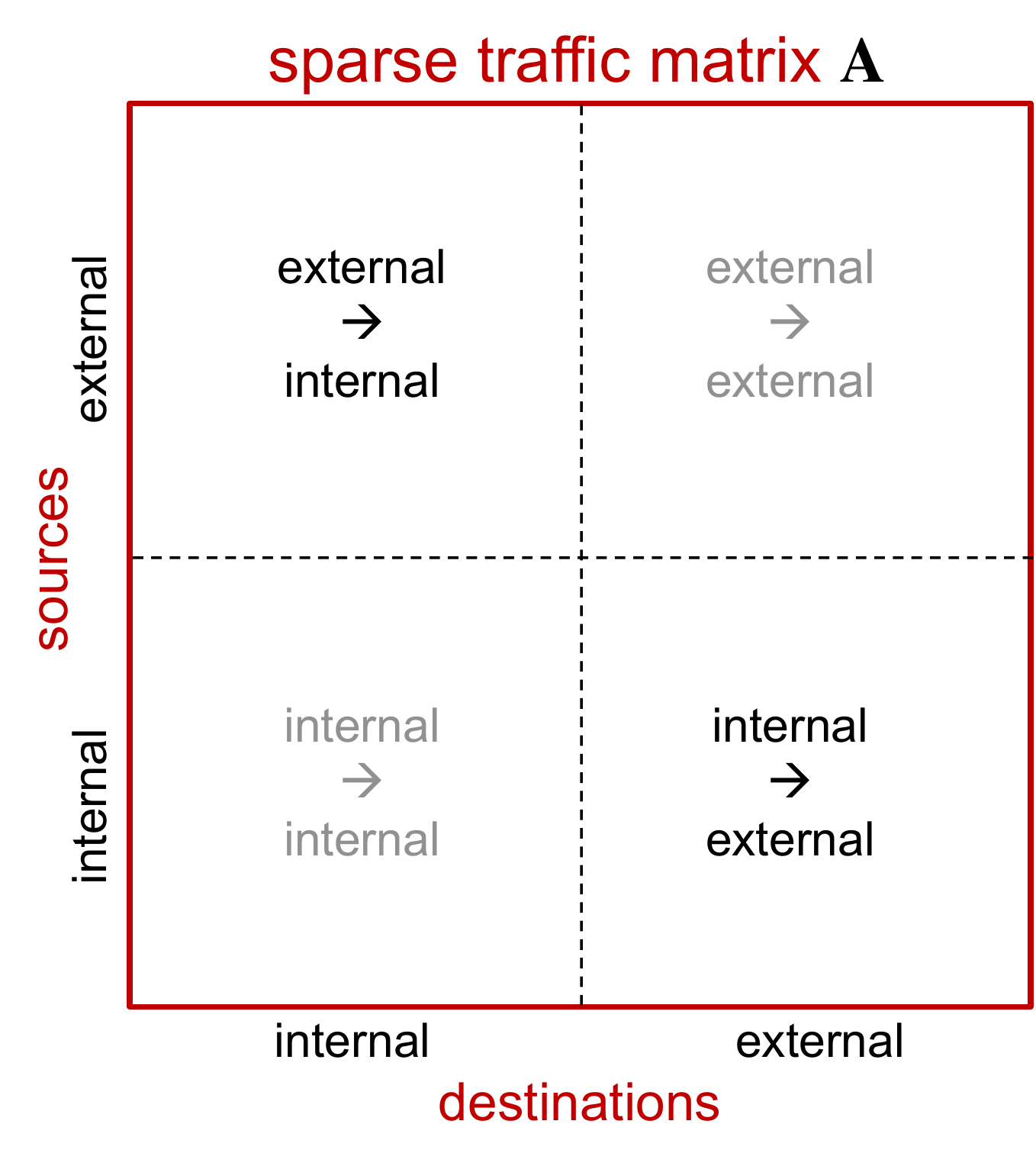}}
      	\caption{{\bf Gateway network traffic matrix.} The traffic matrix of a gateway can be divided into  quadrants separating internal and external traffic.  The gateway will observe the external $\rightarrow$ internal traffic (upper left) and the internal $\rightarrow$ external traffic (lower right).}
      	\label{fig:GatewayTrafficMatrix}
\end{figure}

Network gateways are a common view into the Internet, are widely monitored, and represent a natural observation point on network traffic.  Using such a gateway we were able to collect trillions of packets over several years.  The traffic matrix of a gateway can be partitioned into four quadrants (see Figure~\ref{fig:GatewayTrafficMatrix}).  These quadrants represent different flows between nodes internal and external to the gateway.  A gateway will typically see the external $\rightarrow$ internal and internal $\rightarrow$ external traffic.

\begin{figure}
\center{\includegraphics[width=0.85\columnwidth]{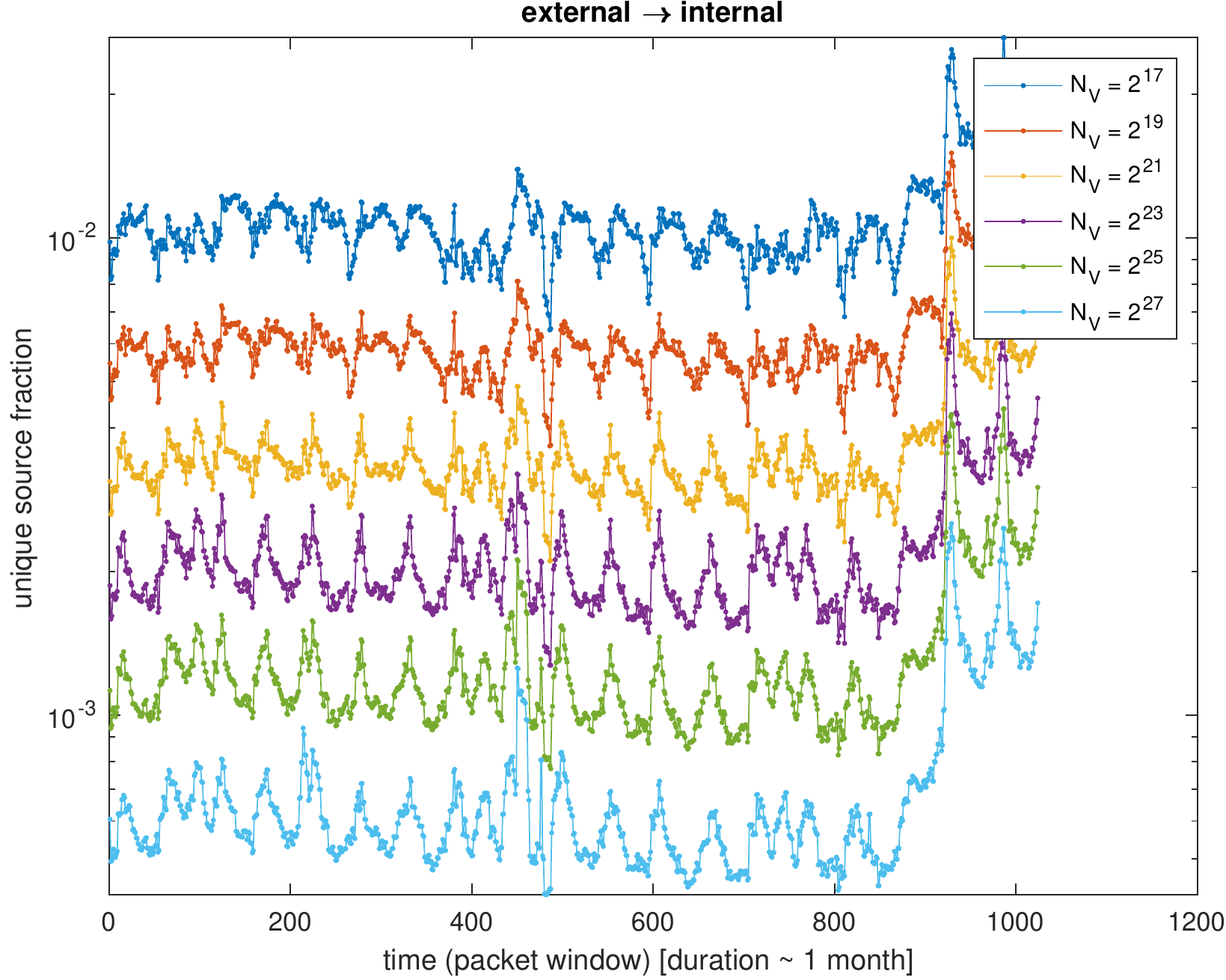}}
\center{\includegraphics[width=0.85\columnwidth]{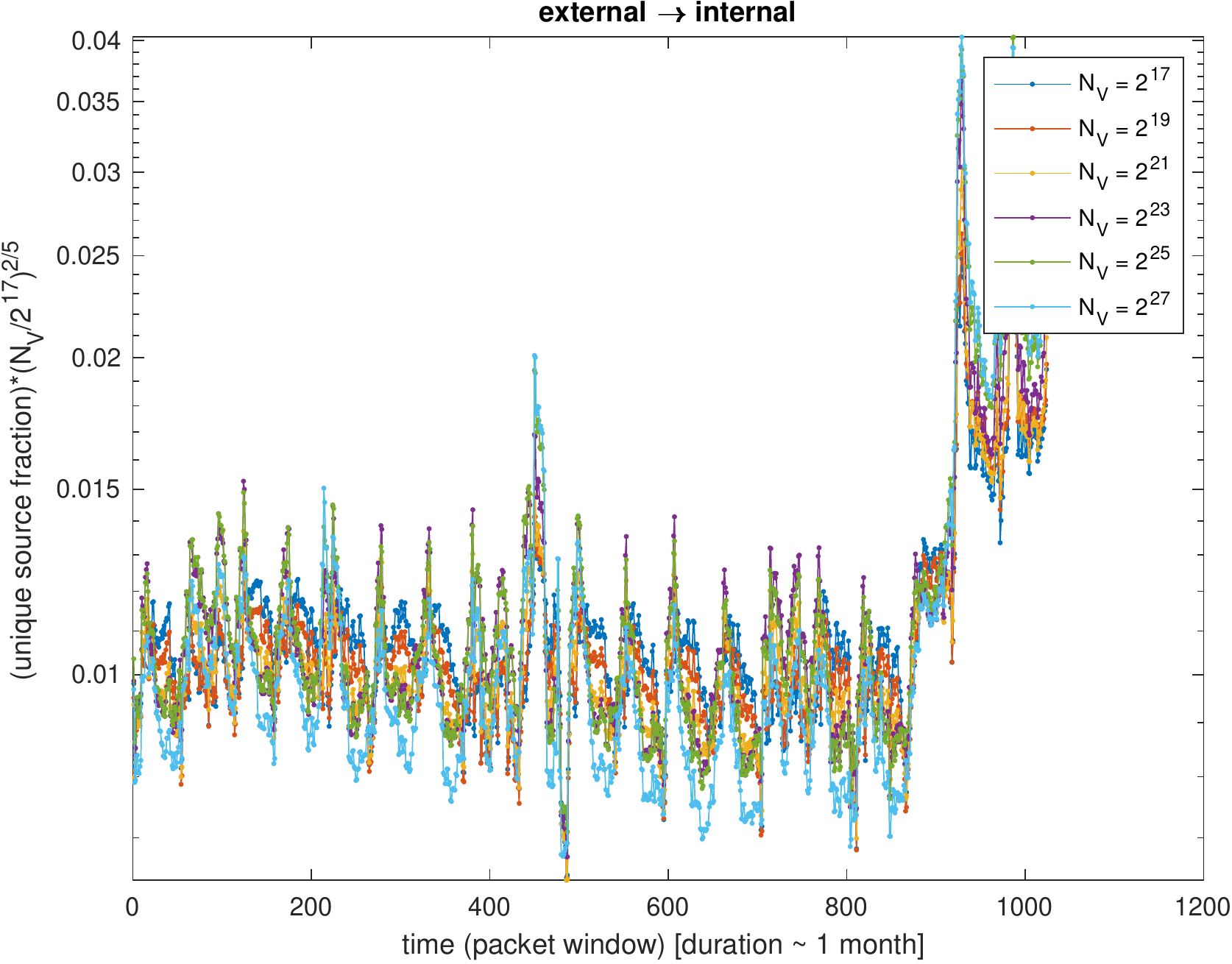}}
      	\caption{{\bf (top) Unique external $\rightarrow$ internal source fraction.}  Average total number of unique sources in a packet window of width $N_V$ measured at each time over a month.  {\bf (bottom) Normalized external $\rightarrow$ internal unique source fraction.}  Scaling (top) data by $(N_V/2^{17})^{2/5}$ aligns the different packet windows, indicating that the number of uniques sources is proportional to $N_V^{1-2/5} = N_V^{3/5}$ (see Table~\ref{tab:ScalingRelations}).}
      	\label{fig:UniqueSourceFraction}
\end{figure}

\begin{table*}[htp]
\caption{Approximate scaling relations.}
\vspace{-0.25cm}
Analysis of network quantities over packet windows $N_V = 2^{17}, \ldots, 2^{27}$ reveals a strong dependence of many of these quantities on $N_V$.  Blank entries indicate that  no simple scaling relation was observed.\center{\includegraphics[width=1.75\columnwidth]{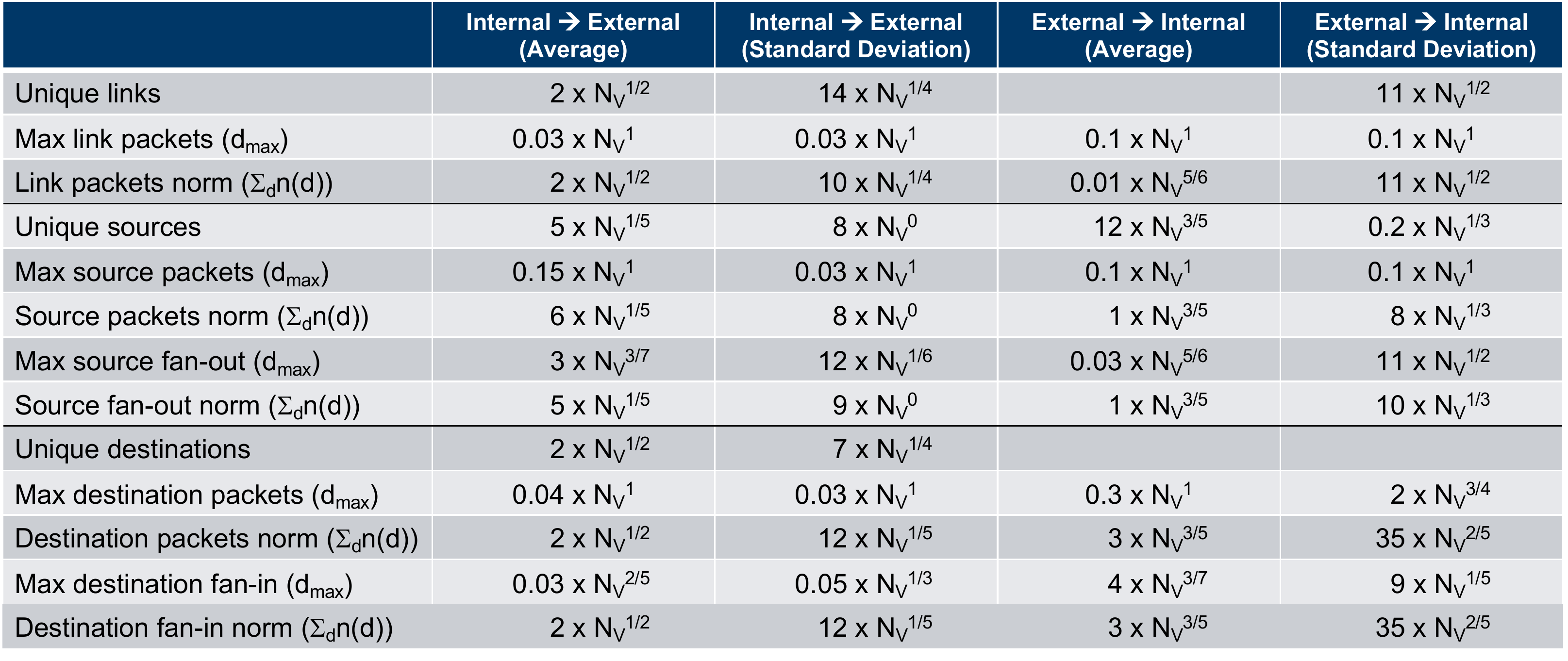}}
\label{tab:ScalingRelations}
\end{table*}%

\section{Results}

Several trillion packet headers have been collected at a gateway over several years.   This work focuses on analyzing one month of anonymized source-destination pairs derived from approximately 100,000,000,000 packet headers.  The contiguous nature of these data allows the exploration of a wide range of packet windows from $N_V = 2^{17}$ (seconds) to $N_V = 2^{27}$ (hours), providing a unique view into how network quantities depend upon time.  Figure~\ref{fig:UniqueSourceFraction} (top) shows the average total number of unique sources in a packet window of width $N_V$ measured at each time over a month normalized by $N_V$.  Figure~\ref{fig:UniqueSourceFraction} (bottom) is the result of scaling the data by $(N_V/2^{17})^{2/5}$ to align the different packet windows, indicating that the number of uniques sources is proportional to $N_V^{1-2/5} = N_V^{3/5}$.  Performing a similar  analysis across many network quantities produced the scaling relations shown in Table~\ref{tab:ScalingRelations}.  These results reveal a strong dependence on these quantities as function of the packet window size $N_V$. To our knowledge, these scaling relations have not been previously reported and represent a new view on the background behavior of network traffic.

\section{Simple Topology Analysis}

\begin{figure*}
\center{\includegraphics[width=1.5\columnwidth]{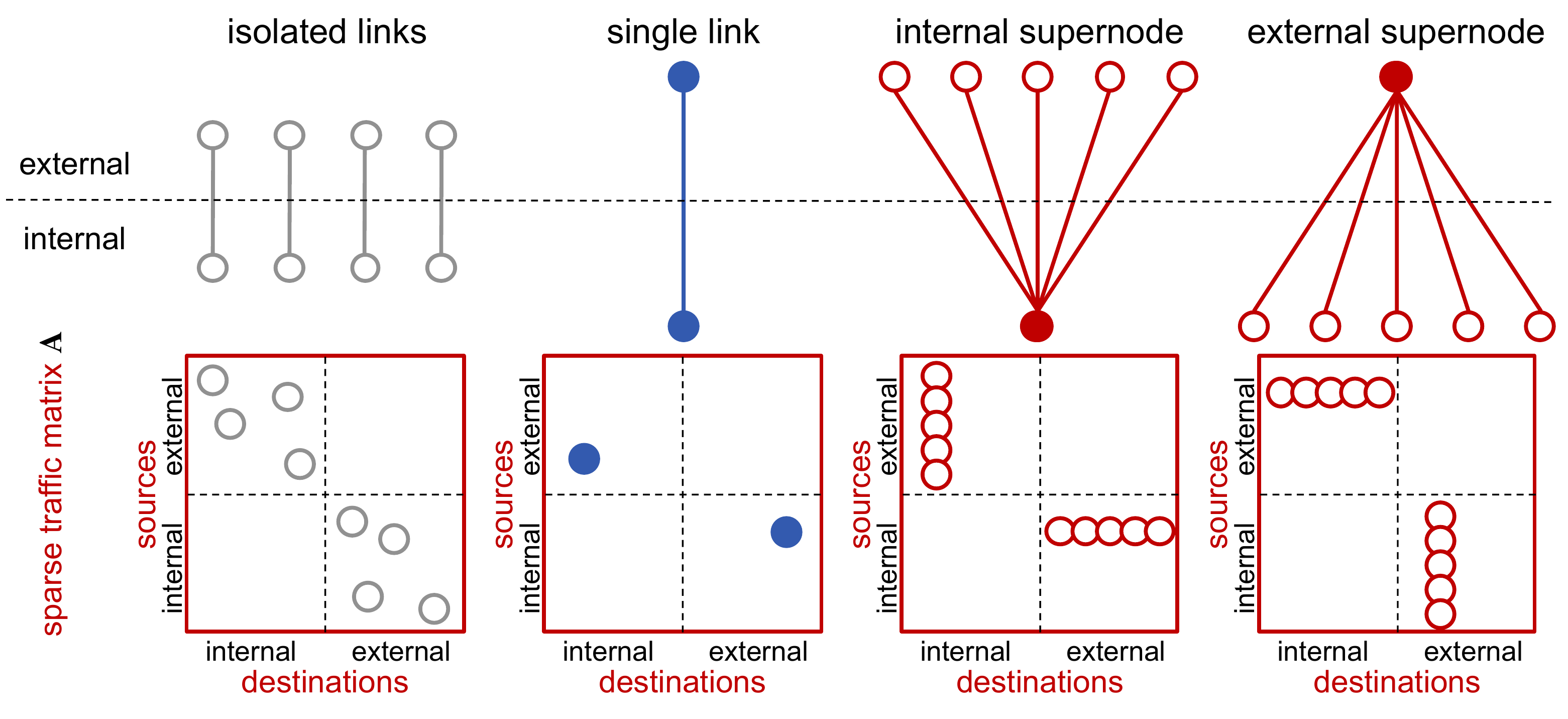}}
      	\caption{{\bf Simple network traffic topologies and traffic matrices.} Isolated links are source-destination pairs that have only one packet. Single link implies all traffic flows between one internal and one external node.   Internal supernode has all traffic occurring between a single internal node and many external nodes. External supernode has all traffic occurring between a single external node and many internal nodes.}
      	\label{fig:TrafficTopologies}
\end{figure*}

Power-law network data are a result of complex network topologies that are an ongoing area of investigation \cite{cao18impact,broido2019scale}. Bounds on the underlying network topologies can be derived by analyzing several simple topologies.  Figure~\ref{fig:TrafficTopologies} shows a schematic of networks and gateway traffic matrices for four simple topologies whose $N_V$ scaling behavior can be readily computed.  Isolated links are source-destination pairs that have only one packet. Single link implies all traffic flows between one internal and one external node.   An internal supernode has all traffic occurring between a single internal node and many external nodes. An external supernode has all traffic occurring between a single external node and many internal nodes.  The resulting scaling relationships are shown in Table~\ref{tab:SimpleScalingRelations}.  In all cases the simple traffic topologies produce relationships that scale as $N_V^0$ (constant) or $N_V^1$ (linear).  The observed relationships in Table~\ref{tab:ScalingRelations} are between these two extremes consistent with observed power-law topologies.  In particular, the observed maximum link, source, and destination packets are consistent with the single link model, suggesting that a fraction of the traffic is on a single link perhaps combining an internal supernode and an external supernode.

\section{Conclusions and Future Work}

We have developed an efficient method for computing a wide variety of streaming network quantities on diverse time scales.  Applying these methods to 100,000,000,000 anonymized source-destination pairs collected at a network gateway reveals many previously unobserved scaling relationships as a function of the packet window  $N_V$.  These observations provide new insights into normal network background traffic.  While the details of these scaling relationships are likely to be specific to the gateway, the number and variety  of relationships suggests that scaling relationships could be observed at other network vantage points.  The observed relationships provide simple, low-dimensional models of the network traffic that could be used for anomaly detection.  These relationships can also assist with feature engineering for the development of AI based traffic categorization.  Deeper theoretical models can also be constrained and tested, and the  observed relationships can provide a useful target for theoreticians.  Future work will expand this analysis to the full several trillion packet data set and other data sets collected at different vantage points, apply these results to anomaly detection problems, and test theoretical models of dynamic streaming networks.

\section*{Acknowledgments}

The authors wish to acknowledge the following individuals for their contributions and support: Bob Bond, K Claffy, Cary Conrad, David Clark, Alan Edelman, Jeff Gottschalk, Tim Kraska, Charles Leiserson, Dave Martinez, Mimi McClure, Steve Rejto, Marc Zissman.

\begin{table*}[htp]
\caption{Simple topology scaling relations.}
\vspace{-0.25cm}
For balanced traffic with equal numbers of packets flowing in each direction, the simple traffic topologies produce relationships that scale as $N_V^0$ (constant) or $N_V^1$ (linear).
\center{\includegraphics[width=1.75\columnwidth]{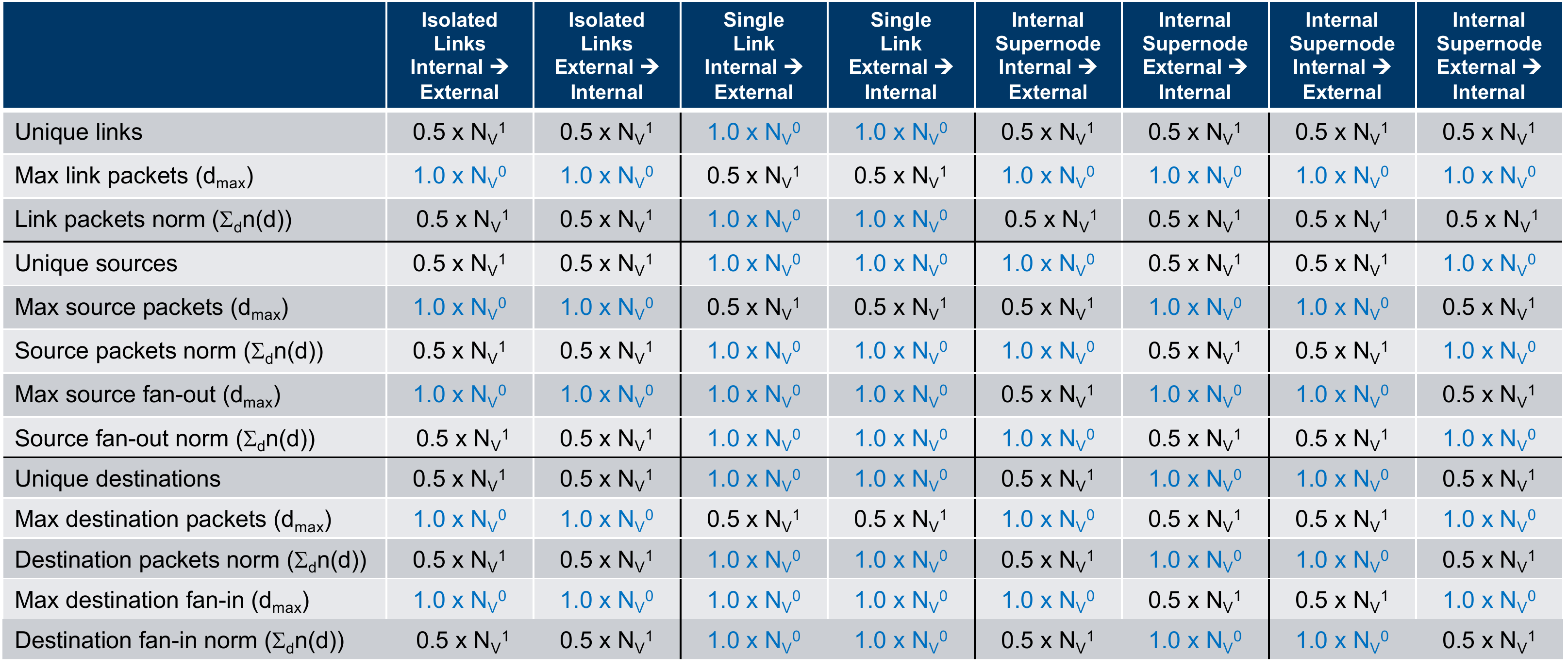}}
\label{tab:SimpleScalingRelations}
\end{table*}%

\bibliographystyle{ieeetr}
\bibliography{StreamingNetworkAnalysis}

\begin{thebibliography}{10}

\bibitem{Cisco2018-2023}
``{\it Cisco Visual Networking Index: Forecast and Trends, 2018–2023}.''
  https://www.cisco.com/c/en/us/solutions/collateral/executive-perspectives/annual-internet-report/white-paper-c11-741490.html.

\bibitem{Bush1945}
V.~Bush, ``{Science The Endless Frontier}.'' Report to the President of the
  United States, 1945.

\bibitem{claffy1999internet}
K.~Claffy, ``Internet tomography,'' {\em Nature, Web Matter}, 1999.

\bibitem{prasad2003bandwidth}
R.~Prasad, C.~Dovrolis, M.~Murray, and K.~Claffy, ``Bandwidth estimation:
  metrics, measurement techniques, and tools,'' {\em IEEE network}, vol.~17,
  no.~6, pp.~27--35, 2003.

\bibitem{kim2008internet}
H.~Kim, K.~C. Claffy, M.~Fomenkov, D.~Barman, M.~Faloutsos, and K.~Lee,
  ``Internet traffic classification demystified: myths, caveats, and the best
  practices,'' in {\em Proceedings of the 2008 ACM CoNEXT conference}, p.~11,
  ACM, 2008.

\bibitem{boguna2009navigability}
M.~Boguna, D.~Krioukov, and K.~C. Claffy, ``Navigability of complex networks,''
  {\em Nature Physics}, vol.~5, no.~1, p.~74, 2009.

\bibitem{claffy2000measuring}
K.~Claffy, ``Measuring the internet,'' {\em IEEE Internet Computing}, vol.~4,
  no.~1, pp.~73--75, 2000.

\bibitem{li2013survey}
B.~Li, J.~Springer, G.~Bebis, and M.~H. Gunes, ``A survey of network flow
  applications,'' {\em Journal of Network and Computer Applications}, vol.~36,
  no.~2, pp.~567--581, 2013.

\bibitem{rabinovich2016measuring}
M.~Rabinovich and M.~Allman, ``Measuring the internet,'' {\em IEEE Internet
  Computing}, vol.~20, no.~4, pp.~6--8, 2016.

\bibitem{ClaffyClark2020}
k.~claffy and D.~Clark, ``Workshop on internet economics (wie 2019) report,''
  {\em SIGCOMM Comput. Commun. Rev.}, vol.~50, p.~53–59, May 2020.

\bibitem{huang2018software}
D.~Huang, A.~Chowdhary, and S.~Pisharody, {\em Software-Defined networking and
  security: from theory to practice}.
\newblock CRC Press, 2018.

\bibitem{lumsdaine2007challenges}
A.~Lumsdaine, D.~Gregor, B.~Hendrickson, and J.~Berry, ``Challenges in parallel
  graph processing,'' {\em Parallel Processing Letters}, vol.~17, no.~01,
  pp.~5--20, 2007.

\bibitem{kolda2009tensor}
T.~G. Kolda and B.~W. Bader, ``Tensor decompositions and applications,'' {\em
  SIAM review}, vol.~51, no.~3, pp.~455--500, 2009.

\bibitem{hilbert2011world}
M.~Hilbert and P.~L{\'o}pez, ``The world's technological capacity to store,
  communicate, and compute information,'' {\em Science}, p.~1200970, 2011.

\bibitem{Kepner2009}
J.~Kepner, {\em Parallel MATLAB for Multicore and Multinode Computers}.
\newblock SIAM, 2009.

\bibitem{kepner2011graph}
J.~Kepner and J.~Gilbert, {\em Graph algorithms in the language of linear
  algebra}.
\newblock SIAM, 2011.

\bibitem{kepner2018mathematics}
J.~Kepner and H.~Jananthan, {\em Mathematics of big data: Spreadsheets,
  databases, matrices, and graphs}.
\newblock MIT Press, 2018.

\bibitem{reuther2018interactive}
A.~{Reuther}, J.~{Kepner}, C.~{Byun}, S.~{Samsi}, W.~{Arcand}, D.~{Bestor},
  B.~{Bergeron}, V.~{Gadepally}, M.~{Houle}, M.~{Hubbell}, M.~{Jones},
  A.~{Klein}, L.~{Milechin}, J.~{Mullen}, A.~{Prout}, A.~{Rosa}, C.~{Yee}, and
  P.~{Michaleas}, ``Interactive supercomputing on 40,000 cores for machine
  learning and data analysis,'' in {\em 2018 IEEE High Performance extreme
  Computing Conference (HPEC)}, pp.~1--6, 2018.

\bibitem{gadepally2018hyperscaling}
V.~{Gadepally}, J.~{Kepner}, L.~{Milechin}, W.~{Arcand}, D.~{Bestor},
  B.~{Bergeron}, C.~{Byun}, M.~{Hubbell}, M.~{Houle}, M.~{Jones},
  P.~{Michaleas}, J.~{Mullen}, A.~{Prout}, A.~{Rosa}, C.~{Yee}, S.~{Samsi}, and
  A.~{Reuther}, ``Hyperscaling internet graph analysis with d4m on the mit
  supercloud,'' in {\em 2018 IEEE High Performance extreme Computing Conference
  (HPEC)}, pp.~1--6, Sep. 2018.

\bibitem{kepner19streaming}
J.~{Kepner}, V.~{Gadepally}, L.~{Milechin}, S.~{Samsi}, W.~{Arcand},
  D.~{Bestor}, W.~{Bergeron}, C.~{Byun}, M.~{Hubbell}, M.~{Houle}, M.~{Jones},
  A.~{Klein}, P.~{Michaleas}, J.~{Mullen}, A.~{Prout}, A.~{Rosa}, C.~{Yee}, and
  A.~{Reuther}, ``Streaming 1.9 billion hypersparse network updates per second
  with d4m,'' in {\em 2019 IEEE High Performance Extreme Computing Conference
  (HPEC)}, pp.~1--6, 2019.

\bibitem{cho2000tr}
K.~Cho, K.~Mitsuya, and A.~Kato, ``Traffic data repository at the wide
  project,'' in {\em Proceedings of USENIX 2000 Annual Technical Conference:
  FREENIX Track}, pp.~263--270, 2000.

\bibitem{borgnat2009seven}
P.~Borgnat, G.~Dewaele, K.~Fukuda, P.~Abry, and K.~Cho, ``Seven years and one
  day: Sketching the evolution of internet traffic,'' in {\em INFOCOM 2009,
  IEEE}, pp.~711--719, IEEE, 2009.

\bibitem{dainotti2012issues}
A.~Dainotti, A.~Pescape, and K.~C. Claffy, ``Issues and future directions in
  traffic classification,'' {\em IEEE network}, vol.~26, no.~1, 2012.

\bibitem{himura2013synoptic}
Y.~Himura, K.~Fukuda, K.~Cho, P.~Borgnat, P.~Abry, and H.~Esaki, ``Synoptic
  graphlet: Bridging the gap between supervised and unsupervised profiling of
  host-level network traffic,'' {\em IEEE/ACM Transactions on Networking},
  vol.~21, no.~4, pp.~1284--1297, 2013.

\bibitem{fontugne2017scaling}
R.~Fontugne, P.~Abry, K.~Fukuda, D.~Veitch, K.~Cho, P.~Borgnat, and H.~Wendt,
  ``Scaling in internet traffic: a 14 year and 3 day longitudinal study, with
  multiscale analyses and random projections,'' {\em IEEE/ACM Transactions on
  Networking (TON)}, vol.~25, no.~4, pp.~2152--2165, 2017.

\bibitem{dhamdhere2018inferring}
A.~Dhamdhere, D.~D. Clark, A.~Gamero-Garrido, M.~Luckie, R.~K. Mok, G.~Akiwate,
  K.~Gogia, V.~Bajpai, A.~C. Snoeren, and K.~Claffy, ``Inferring persistent
  interdomain congestion,'' in {\em Proceedings of the 2018 Conference of the
  ACM Special Interest Group on Data Communication}, pp.~1--15, ACM, 2018.

\bibitem{kepner16mathematical}
J.~{Kepner}, P.~{Aaltonen}, D.~{Bader}, A.~{Buluç}, F.~{Franchetti},
  J.~{Gilbert}, D.~{Hutchison}, M.~{Kumar}, A.~{Lumsdaine}, H.~{Meyerhenke},
  S.~{McMillan}, C.~{Yang}, J.~D. {Owens}, M.~{Zalewski}, T.~{Mattson}, and
  J.~{Moreira}, ``Mathematical foundations of the graphblas,'' in {\em 2016
  IEEE High Performance Extreme Computing Conference (HPEC)}, pp.~1--9, 2016.

\bibitem{buluc17design}
A.~{Buluc}, T.~{Mattson}, S.~{McMillan}, J.~{Moreira}, and C.~{Yang}, ``Design
  of the graphblas api for c,'' in {\em 2017 IEEE International Parallel and
  Distributed Processing Symposium Workshops (IPDPSW)}, pp.~643--652, 2017.

\bibitem{davis18algorithm}
T.~A. Davis, ``Algorithm 1000: Suitesparse:graphblas: Graph algorithms in the
  language of sparse linear algebra,'' {\em ACM Trans. Math. Softw.}, vol.~45,
  Dec. 2019.

\bibitem{kepner202075}
J.~Kepner, T.~Davis, C.~Byun, W.~Arcand, D.~Bestor, W.~Bergeron, V.~Gadepally,
  M.~Hubbell, M.~Houle, M.~Jones, A.~Klein, P.~Michaleas, L.~Milechin,
  J.~Mullen, A.~Prout, A.~Rosa, S.~Samsi, C.~Yee, and A.~Reuther,
  ``75,000,000,000 streaming inserts/second using hierarchical hypersparse
  graphblas matrices,'' {\em IPDPSW GrAPL}, 2020.

\bibitem{elmore2015demonstration}
A.~J. Elmore, J.~Duggan, M.~Stonebraker, M.~Balazinska, U.~Cetintemel,
  V.~Gadepally, J.~Heer, B.~Howe, J.~Kepner, T.~Kraska, {\em et~al.}, ``A
  demonstration of the bigdawg polystore system,'' {\em Proceedings of the VLDB
  Endowment}, vol.~8, no.~12, p.~1908, 2015.

\bibitem{kraska18case}
T.~Kraska, A.~Beutel, E.~H. Chi, J.~Dean, and N.~Polyzotis, ``The case for
  learned index structures,'' in {\em Proceedings of the 2018 International
  Conference on Management of Data}, SIGMOD 18, (New York, NY, USA),
  p.~489â504, Association for Computing Machinery, 2018.

\bibitem{do20classifying}
E.~H. {Do} and V.~N. {Gadepally}, ``Classifying anomalies for network
  security,'' in {\em ICASSP 2020 - 2020 IEEE International Conference on
  Acoustics, Speech and Signal Processing (ICASSP)}, pp.~2907--2911, 2020.

\bibitem{soule2004identify}
A.~Soule, A.~Nucci, R.~Cruz, E.~Leonardi, and N.~Taft, ``How to identify and
  estimate the largest traffic matrix elements in a dynamic environment,'' in
  {\em ACM SIGMETRICS Performance Evaluation Review}, vol.~32, pp.~73--84, ACM,
  2004.

\bibitem{zhang2005estimating}
Y.~Zhang, M.~Roughan, C.~Lund, and D.~L. Donoho, ``Estimating point-to-point
  and point-to-multipoint traffic matrices: an information-theoretic
  approach,'' {\em IEEE/ACM Transactions on Networking (TON)}, vol.~13, no.~5,
  pp.~947--960, 2005.

\bibitem{mucha2010community}
P.~J. Mucha, T.~Richardson, K.~Macon, M.~A. Porter, and J.-P. Onnela,
  ``Community structure in time-dependent, multiscale, and multiplex
  networks,'' {\em science}, vol.~328, no.~5980, pp.~876--878, 2010.

\bibitem{tune2013internet}
P.~Tune, M.~Roughan, H.~Haddadi, and O.~Bonaventure, ``Internet traffic
  matrices: A primer,'' {\em Recent Advances in Networking}, vol.~1, pp.~1--56,
  2013.

\bibitem{karvanen2003measuring}
J.~Karvanen and A.~Cichocki, ``Measuring sparseness of noisy signals,'' in {\em
  4th International Symposium on Independent Component Analysis and Blind
  Signal Separation}, pp.~125--130, 2003.

\bibitem{clauset2009power}
A.~Clauset, C.~R. Shalizi, and M.~E. Newman, ``Power-law distributions in
  empirical data,'' {\em SIAM review}, vol.~51, no.~4, pp.~661--703, 2009.

\bibitem{barabasi2016network}
A.-L. Barab{\'a}si {\em et~al.}, {\em Network science}.
\newblock Cambridge university press, 2016.

\bibitem{leland1994self}
W.~E. Leland, M.~S. Taqqu, W.~Willinger, and D.~V. Wilson, ``On the
  self-similar nature of ethernet traffic (extended version),'' {\em IEEE/ACM
  Transactions on Networking (ToN)}, vol.~2, no.~1, pp.~1--15, 1994.

\bibitem{faloutsos1999power}
M.~Faloutsos, P.~Faloutsos, and C.~Faloutsos, ``On power-law relationships of
  the internet topology,'' in {\em ACM SIGCOMM computer communication review},
  vol.~29-4, pp.~251--262, ACM, 1999.

\bibitem{albert1999internet}
R.~Albert, H.~Jeong, and A.-L. Barab{\'a}si, ``Internet: Diameter of the
  world-wide web,'' {\em Nature}, vol.~401, no.~6749, p.~130, 1999.

\bibitem{barabasi1999emergence}
A.-L. Barab{\'a}si and R.~Albert, ``Emergence of scaling in random networks,''
  {\em Science}, vol.~286, no.~5439, pp.~509--512, 1999.

\bibitem{adamic2000power}
L.~A. Adamic and B.~A. Huberman, ``Power-law distribution of the world wide
  web,'' {\em science}, vol.~287, no.~5461, pp.~2115--2115, 2000.

\bibitem{barabasi2009scale}
A.-L. Barab{\'a}si, ``Scale-free networks: a decade and beyond,'' {\em
  science}, vol.~325, no.~5939, pp.~412--413, 2009.

\bibitem{mahanti2013tale}
A.~Mahanti, N.~Carlsson, A.~Mahanti, M.~Arlitt, and C.~Williamson, ``A tale of
  the tails: Power-laws in internet measurements,'' {\em IEEE Network},
  vol.~27, no.~1, pp.~59--64, 2013.

\bibitem{cao18impact}
Z.~Cao, Z.~He, and N.~F. Johnson, ``Impact on the topology of power-law
  networks from anisotropic and localized access to information,'' {\em Phys.
  Rev. E}, vol.~98, p.~042307, Oct 2018.

\bibitem{broido2019scale}
A.~D. Broido and A.~Clauset, ``Scale-free networks are rare,'' {\em Nature
  communications}, vol.~10, no.~1, pp.~1--10, 2019.

\end{thebibliography}

\appendices
\setcounter{equation}{0}
\renewcommand{\theequation}{\thesection\arabic{equation}}

\end{document}